\documentclass{article}

\usepackage{arxiv}

\usepackage[utf8]{inputenc} 
\usepackage[T1]{fontenc}    
\usepackage{hyperref}       
\usepackage{url}            
\usepackage{booktabs}       
\usepackage{amsfonts}       
\usepackage{nicefrac}       
\usepackage{microtype}      
\usepackage{lipsum}
\usepackage{graphicx}
\graphicspath{ {./images/} }

\usepackage{amsmath}
\usepackage{algorithm}
\usepackage{algorithmic}
\usepackage{enumerate}
\usepackage{xcolor}

\newcommand\Blue[1]{\textcolor{blue}{#1}}

\title{Interactive volume illumination of slice-based ray casting}

\author{
 Dening Luo \\
  College of Computer Science, SiChuan University\\
  No. 24 South Section 1, Yihuan Road, Chengdu Sichuan, China, 610065\\
  \texttt{loening@stu.scu.edu.cn} \\
}

\begin{document}
\maketitle
\begin{abstract}
Volume rendering always plays an important role in the field of medical imaging and industrial design. In recent years, the realistic and interactive volume rendering of the global illumination can improve the perception of shape and depth of volumetric datasets. In this paper, a novel and flexible performance method of slice-based ray casting is proposed to implement the volume illumination effects, such as volume shadow and other scattering effects. This benefits from the slice-based illumination attenuation buffers of the whole geometry slices at the viewpoint of the light source and the high-efficiency shadow or scattering coefficient calculation per sample in ray casting. These tests show the method can obtain much better volume illumination effects and more scalable performance in contrast to the local volume illumination in ray casting volume rendering or other similar slice-based global volume illumination.
\end{abstract}


\section{Introduction}
In the numerous visualization fields of volumetric datasets, medical imaging and industrial design applications are the two more active domains of exploration. Computed tomography (CT)/magnetic resonance imaging (MRI) datasets are visualized intuitively with full 3D effects in medical imaging. In industrial design, some intermediate results are visually simulated from the data flow or structural data analysis. Therefore, the visualization of volumetric datasets has produced great practical value.

With the development of data visualization techniques, plenty of volume rendering methods have also been widely proposed. Some have been used in practical products, such as slice-based rendering, ray casting volume rendering, etc. Different ways have both strengths and weaknesses when applied to various specific applications. Fortunately, volume rendering methods are still being studied, optimized, and improved to benefit better real-world applications, among which volume illumination mainly addresses problems of the visual perception of depth and spatial structure.

Usually, volume illumination requires a complicated procedure of illumination calculation or data processing. Particularly, when the data size and algorithm complexity increase tremendously, more and more significant challenges are presented for interactive volume rendering. Nowadays, figuring out how to better utilize GPU and novel rendering techniques to resolve performance and effect issues has also become a significant challenge.

In this paper, the interactive volume illumination of slice-based ray casting is proposed to obtain the scalable performance and the outstanding volume illumination effects to tackle the challenges mentioned above. The specific and promising contribution are as follows:

\begin{itemize}

\item This is the first time to combine the two methods based on slice-based volume rendering and ray casting volume rendering to render the global volume illumination of volumetric datasets. In this way, the advantages of the respective algorithms can be used to improve the efficiency and effect. The slice-based is easy to simplify the calculation of the volumetric illumination process, while the ray casting method can efficiently render volumetric datasets. 
 
\item The idea of having illumination attenuation slices from the light source that can be used in volume ray casting for calculating the shadow or some scattering effects. Meanwhile, varying the number of slices and per-slice image resolution allows for flexible performance in volume illumination rendering. 

\item The scattering coefficient of two types, shell and cone, approximates the calculation of volume illumination integration in volume ray casting. The use of illumination attenuation slices with the use of a scattering coefficient can remove any problems from alpha-blending in contrast to the slice-based volume illumination.

\end{itemize}

\section{Related work}
Volume visualization has always been one of the most interesting areas in scientific visualization \cite{beyer2018gpu}. Specifically, it is about a method of extracting meaningful information from volumetric datasets using interactive graphics and imaging techniques. It is concerned with the representation, manipulation, and rendering of volumetric datasets. Meanwhile, with the development of display devices, volumetric datasets from numerous simulation and sampling devices such as MRI, CT, positron emission tomography (PET), ultrasonic imaging, confocal microscopy, supercomputer simulations, geometric models, laser scanners, depth images estimated by stereo disparity, satellite imaging, and sonar can be efficiently visualized and demonstrated on web \cite{MK:16}, mobile \cite{hachaj2014real, noguera2015mobile} or virtual reality \cite{HW:16} platforms or devices.

Volume rendering has been developed for many years, and a large number of interactive rendering techniques have been proposed and practically applied \cite{JS:14}. Currently, direct volume rendering has become more important due to effectiveness compared with indirect volume rendering, which is a method to reconstruct geometry from volumetric datasets \cite{EM:19}. Therefore, different methods are proposed to achieve direct volume rendering such as slice-based \cite{BS:05, Milan}, ray casting \cite{kruger2003acceleration, SSK:05}, shear-warp \cite{LL:94} and splatting \cite{LH:91}. In summary, a clear and fast representation of the 3D structures and internal details of volumetric datasets is the key task of volume visualization. 

The 3D texture slicing volume rendering is the simplest and fastest slice-based approach on GPU. It approximates the volume-density function by slicing the datasets in front-to-back or back-to-front order and then blends the slices using hardware-supported blending. The view-aligned 3D texture slicing \cite{EKE:01, FR:04} is a classic method, and it takes advantage of functionalities implemented on graphics hardware like rasterization, texturing and blending. The half-angle slicing for volume shadow \cite{KPHE:02, EHKRW:06} is also a global illumination technique. The half-angle slicing requires a large number of passes; thus, the draw calls are extensive to result in high-performance overhead. Meanwhile, various composition schemes \cite{ZE:11} are used with particular purposes, including first, average, maximum intensity projection, accumulation, etc.

Ray casting volume rendering is easily implemented on a single-pass pass GPU. Meanwhile, the normal at the sampling point in ray casting can be estimated to carry out the lighting calculation. Although local volume illumination is adequate for shading surfaces, it does not provide sufficient lighting characteristics for the visual appearance. The global volume illumination \cite{SMP:11, zhang2013lighting} presents detailed lighting characteristics in contrast to the local volume illumination. Perceptually motivated techniques for the visualization add additional insights when displaying the volumetric datasets since global illumination techniques can obviously improve spatial depth and shape cues, thus provide better perception \cite{ZY:13, PB:16}. Subsequently, based on the directional occlusion approach by Schott et al. \cite{schott2009directional}, later approaches \cite{vsolteszova2010multidirectional, patel2013instant} used convolution-based illumination where on the directional viewing and illumination buffers can be computed in the same pass using multiple render targets.

The transfer function (TF) \cite{LKGHHY:16, ME:17} for volume rendering is a central topic in scientific visualization and is used to present more details of volumetric datasets, for example implementing volume classification. Besides, the center topics in scientific visualization are parallel volume rendering \cite{UJ:17, BHP:15}, and learning-based volume rendering \cite{WC:19} in the future. Considering the better perception of volumetric datasets and the flexibility and scalability of the growing data, this paper's slice-based ray casting algorithm is proposed to demand the increasing practical applications of interactive volume rendering.

\section{Algorithm}
Volumetric datasets are usually a 3D array of volume elements or voxels, so volume rendering is the reconstruction process of displaying each point in a volume. Voxels can represent various physical characteristics, such as density, temperature, velocity, and pressure. Typically, the volume datasets store densities obtained using a cross-sectional imaging modality such as CT or MRI scans. A 3D texture, which is an array of 2D textures, is obtained by accumulating these 2D slices along the slicing axis.

For a long time, volume rendering implementations are almost exclusively based on slice-based methods where axis-aligned or view-aligned texture slicing is blended to approximate the volume rendering integration. Volume rendering integration is a complex process and is difficult to compute because the complete equation of light transportation is computationally intensive. So, in most practical applications, simplified models are often used. The conventional models include absorption only, emission only, emission-absorption model, single scattering and shadowing, and multiple scattering. 

Light will attenuate as the medium travels. For slice-based illumination attenuation, the illumination intensity of the slice attenuates proportionally with the distance to the light source, and the simulation is performed slice by slice according to the blend function. We can obtain the illumination attenuation per voxels along the light direction and use the idea of storing illumination attenuation slices to calculate the illumination of volumetric datasets in ray casting. Correctly, on the one hand, the illumination attenuation per voxel of the whole volume under the light source can be stored in a buffer; on the other side, the ray casting can traverse the entire volumetric datasets to reconstruct volume information per pixel in image space. Therefore, combined with the advantages of both, we can achieve some better volume illumination effects.

\begin{figure}[htb]
  \centering
  \includegraphics[width=1.0\linewidth]{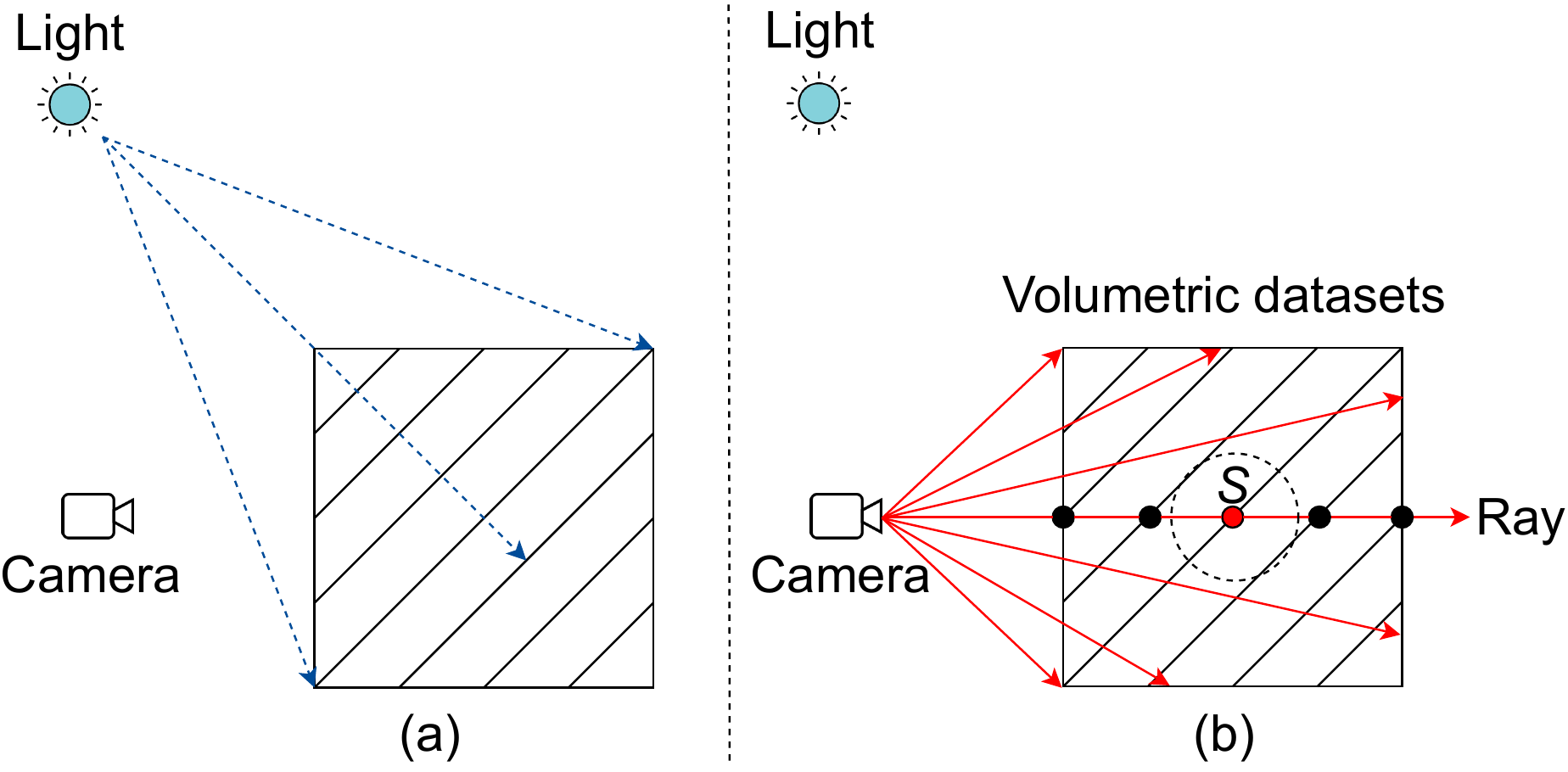}
  \parbox[t]{.9\columnwidth}{\relax
           (a) Render the whole geometry slices of volumetric datasets into $\Blue{LightBuffer}$ slice by slice at the viewpoint of the light source. (b) Reconstruct the 3D volume using volume ray casting and compute the volume illumination effects using the $\Blue{LightBuffer}$ at the viewpoint of the camera.
           }
  \caption{\label{fig:Overview}
           The rendering process of slice-based ray casting.}
\end{figure}

As shown in \textbf{Figure~\ref{fig:Overview}}, the slice-based ray casting first renders the whole geometry slices of volumetric datasets into $\Blue{LightBuffer}$ slice by slice at the view viewpoint of the light source. And then reconstruct the 3D volume using volume ray casting and compute the volume illumination effects using the $\Blue{LightBuffer}$ at the viewpoint of the camera. Correctly, the following three approaches are used to calculate the illumination coefficient of each voxel. The procedure which directly calculates the attenuation coefficient of the sample point in ray casting is the simplest method to compute volume shadow. The shell and cone scattering distributions are the approximation approaches to the global illumination effects of volumetric datasets.

\subsection{Illumination calculation}

\subsubsection{Slice-based illumination attenuation }
\begin{figure}[htb]
  \centering
  \includegraphics[width=.8\linewidth]{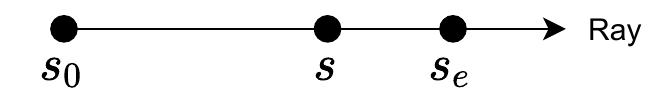}
  \caption{\label{fig:Ray}
           A ray along the light direction in volume illumination.}
\end{figure}

Volume illumination can present the spatial perception of volumetric datasets, and it is different from the surface/local illumination mode. Usually, the volume illumination mode \cite{rostamzadeh2013comparison} in its differential form is solved by integrating (see Equation \ref{eq:VolumeIlluminationMode}) along the light direction from the starting point $s_{0}$ to the endpoint $s_{e}$ (see \textbf{Figure~\ref{fig:Ray}}). The light intensity $I_{s_{e}}$ at $s_{e}$ consists of two parts, which one is the attenuation intensity along the ray, and the other is the integrated contributions of all points along the ray.

\begin{equation}
\left\{
    \begin{array}{lr}
    I_{s_{e}} = I_{s_{0}}e^{-\tau (s_{0},s_{e})}+\int_{s_{0}}^{s_{e}}q(s)e^{-\tau (s,s_{e})}ds \\
    \tau (s,s_{e})=\int_{s}^{s_{e}}\kappa (t)dt \\
    \end{array}
\right.
\label{eq:VolumeIlluminationMode}
\end{equation}
with $I_{s_{0}}$ is initial intensity at $s_{0}$, $e^{-\tau (s_{0},s_{e})}$ is the attenuation along the ray, optical depth $\tau$, optical properties $\kappa$ (absorption coefficient) and $q$ (source term describing emission). 

In most practical applications, simplified models are often used because the complete equation of light transportation is computationally intensive. The emission-absorption model, which is most common in volume rendering, is used. For the emission-absorption model, the accumulated color and opacity are computed according to Equation \ref{eq:emission-absorption}, where $C_{i}$ and $A_{i}$ are the color and opacity assigned by the transfer function to the data value at sample $i$. Opacity $A_{i}$ approximates the absorption, and opacity-weighted color $C_{i}$ approximates the emission and the absorption along the ray segment between samples $i$ and $i+1$.

\begin{equation}
\left\{
    \begin{array}{lr}
    C=\sum_{i=1}^{n}C_{i}\prod_{j=1}^{i-1}(1-A_{j}) \\
    A=1 - \prod_{j=1}^{n}(1-A_{j}) \\
    \end{array}
\right.
\label{eq:emission-absorption}
\end{equation}

For the iterative computations of the discretization volume integration, the blend function is different for either front-to-back or back-to-front. As shown in Equation \ref{eq:FrontToBack} and Equation \ref{eq:BackToFront}, respectively, the front-to-back means rendering proceeds in front-to-back order to the eye, and the back-to-front means in back-to-front order to the eye. Variables with subscript $_{src}$ ( as for "source") describe quantities introduced as inputs from the optical properties of the data set (e.g., through a transfer function). In contrast, variables with subscript $_{dst}$ (as for "destination") describe output quantities that hold accumulated colors and opacities.

\begin{equation}
\left\{
    \begin{array}{lr}
    C_{dst}\leftarrow C_{dst}+(1-\alpha _{dst})C_{src} \\
    \alpha _{dst}\leftarrow \alpha_{dst}+(1-\alpha _{dst})\alpha _{src} \\
    \end{array}
\right.
\label{eq:FrontToBack}
\end{equation}

\begin{equation}
C_{dst}\leftarrow (1-\alpha _{src})C_{dst}+C_{src}
\label{eq:BackToFront}
\end{equation}

\subsubsection{Extinction-based illumination coefficient}
Direct volume rendering integration (Equation \ref{eq:DVRIntegral}) \cite{max1995optical} can not be solved analytically without making some confining assumptions and consequently needs to be approximated. A slightly coarse approximation, the classical $\alpha$-blending, is the commonly used development of the original extinction coefficient into a Taylor series where only the first two elements are considered. A closer approximation based on the initial extinction coefficient (Equation \ref{eq:Discretization}) supplies a more accurate evaluation of the volume rendering integration on programmable GPUs since the advantage of integrating over the exponential extinction coefficient is a summation in contrast to the product of $\alpha$-blending.

In Equation \ref{eq:DVRIntegral}, a ray from $s_{0}=0$ at the back of the volume to $s_e$ at the eye position is considered. The extinction coefficient is indicated by $\tau(s)$, and $E(s)$ is the light reflected or emitted by a volume sample at $s$. 

\begin{equation}
I_{s_{e}}=\int_{0}^{s_e}E(s)\tau (s)e^{-\int_{0}^{s_e}\tau (t)dt}ds
\label{eq:DVRIntegral}
\end{equation}

In the discretization of Equation \ref{eq:Discretization} using a step size ($\Delta t$) along the ray, instead of performing a Taylor series expansion and simplification of the extinction term and the original exponential extinction coefficient can be retained. The formulation is sufficiently simple and can easily be implemented on programming GPUs. The additive property of $\tau$ allows for the summation of the samples in a shader in arbitrary order. The basic premise is that any light occlusion and shadowing effects arise from the attenuation of light traveling or being scattered through the volume along a ray or within some specific region. Therefore, any light attenuation stems from some extinction factor $e^{-\sum_{j}{\Delta t}\tau _{j}}$ where the sum $\sum_{j}{\Delta t}\tau _{j}$ must be taken over a ray or region of the volume.

\begin{equation}
I_{s_{e}}\approx \sum_{i=0}^{s_{e}/\Delta t}E_{i}{\Delta t}\tau _{i}e^{-\sum_{j=i}^{s_{e}/\Delta t}{\Delta t}\tau _{j}}
\label{eq:Discretization}
\end{equation}
where the reflected and emitted light $E_i$ is typically replaced by a voxel color modulated by a simple lighting model.

The dense ray calculation of each sample for the light source is costly, mainly referring to light scattering. So, we consider the illumination attenuation of each voxel and compute each sample's attenuation coefficient to simplify the volume illumination calculation in Equation \ref{eq:AttenuationCoefficient}. Moreover, we use the neighborhood sample to yield the necessary extinction of the light on its way. Many scattering lights are estimated by distance function and neighborhood voxels. A more sparsely occluded neighborhood is an indicator that more light is scattered to the sample, thus supporting visually plausible soft shadow borders.

\begin{equation}
I_{s_{e}}\approx \sum_{i=0}^{s_{e}/\Delta t}A_{i}E_{i}{\Delta t}
\label{eq:AttenuationCoefficient}
\end{equation}
where the attenuation coefficient $A_i$ is the weight of each sample point for volume illumination.

\subsection{Slice-based ray casting}
\subsubsection{Build illumination attenuation buffer}
Volume illumination is often a complex and expensive process since illumination calculation needs to trace the propagation of large amounts of light and interact with the scene. The processes of complex illumination calculations are usually simplified to meet the real-time application. Therefore, the paper cleverly uses slice-based illumination attenuation to calculate the propagation of light within a volume. It is a scalable and effective way to calculate the illumination process of volumetric datasets since varying the slice distances and per-slice image resolution allows for flexible performance. 

\begin{figure*}[htb]
  \centering
  \includegraphics[width=1.0\linewidth]{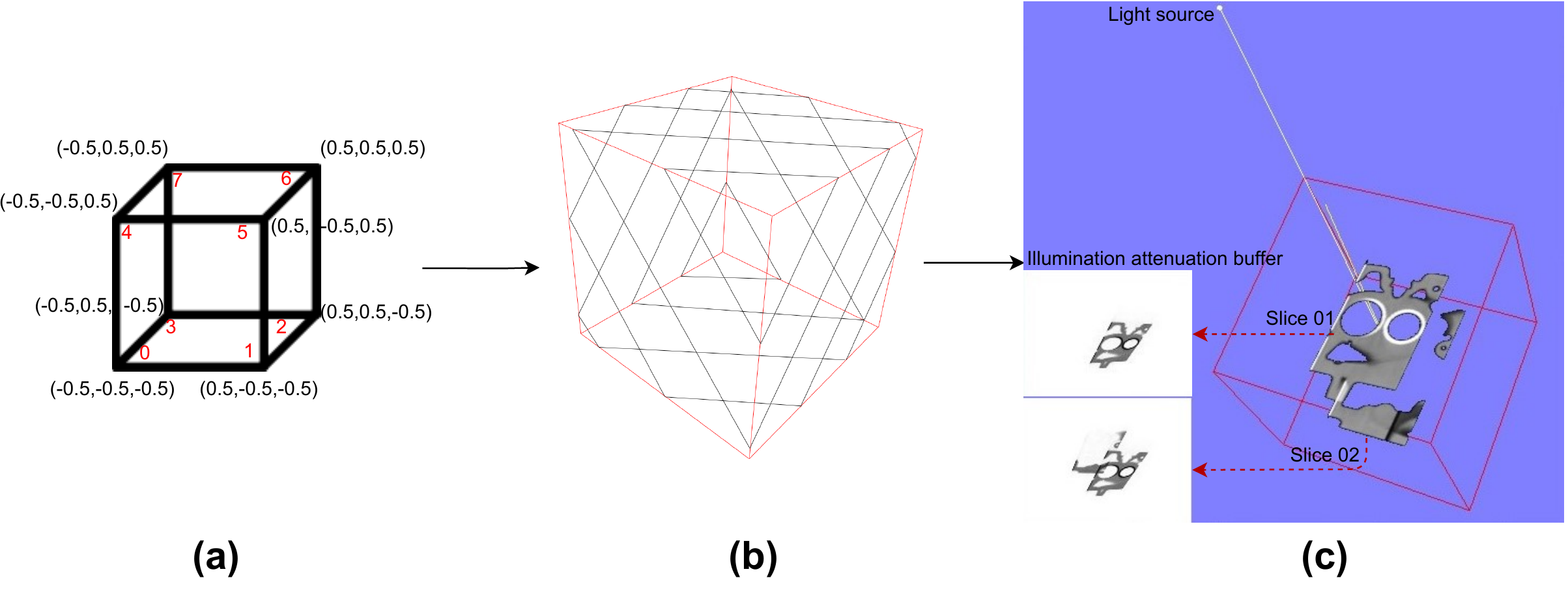}
           
  %
  \caption{\label{fig:LightBuffer}
           The $\Blue{LightBuffer}$ at the viewpoint of the light source. (a) A unit cube; (b) The geometry slices; (c) An example of illumination attenuation slices.}
\end{figure*}

The algorithm \ref{code:LightBuffer} shows the process of building illumination attenuation buffer slice by slice. The primary method is to render each slice of the whole geometry slices of the volumetric datasets into illumination attenuation array at the viewpoint of the light source and accumulate the illumination attenuation buffer slice by slice. The illumination intensity of the current slice is only determined by multiplying the sample color opacity $\alpha$ by illumination color. Each slice color is determined by blending the illumination attenuation of previous slices into the frame buffer in back-to-front. For illumination attenuation, the illumination intensity of the slice attenuates proportionally under the light source, as shown in \textbf{Figure~\ref{fig:LightBuffer}}. The second layer slice is the blended illumination intensity of the first layer slice. The volume shadow is calculated by the illumination attenuation slice by slice.

\begin{algorithm}[h]
\caption{Build illumination attenuation buffer slice by slice}
\label{code:LightBuffer} 
\begin{algorithmic}[1]
\STATE Generate the whole geometry slices;
\STATE Initialize $\Blue{LightBuffer}$;
\STATE Render all slices at the viewpoint of light source;
\FOR{each slice}
\FOR{each sample}
\STATE Evaluate sample opacity $\alpha$;
\STATE Multiply $\alpha$ by illumination color;
\STATE Blend slice into the frame buffer in back-to-front;
\ENDFOR
\STATE Read the frame buffer image and write into $\Blue{LightBuffer}$;
\ENDFOR
\end{algorithmic}
\end{algorithm}

For the illumination attenuation buffer, the volumetric datasets need to be sliced into geometry slices and rendered slice by slice at the viewpoint of the light source. The whole geometry slices are generated by the intersections of a unit cube and the planes which are perpendicular to the light direction. There are roughly the following steps to complete the process.

\begin{enumerate}
  \item Calculate the min/max distance of unit cube vertices by doing a dot product of each unit cube vertex $V[i]$ with the light direction vector $L$;
  \item Calculate all the possible intersections parameter $\lambda$ of the plane perpendicular to the light direction with all edges of the unit cube going from the nearest to farthest vertex, using min/max distances from step 1;
  \item Find the intersection points of each slice using the intersection parameter $\lambda$ (from step 2) to move in the light direction; Three to six intersection vertices will be generated;
  \item Store the intersection points in the specified order to generate triangular primitives;
  \item Update geometry slices.
\end{enumerate} 

The minimum distance is set as the first slice position, and the position increment is the substrate the max and min distances divided by the total number of slices. The min/max distance and the total number of slices can be calculated as shown Equation \ref{eq:Max/Min}. Meanwhile, several parameters and the sample point can help to calculate the correct slice index.

\begin{equation}
\left\{\begin{matrix}
min=\underset{0\leq i\leq 7}{min}\left \{L\cdot V[i]  \right \}\\ 
max=\underset{0\leq i\leq 7}{max}\left \{L\cdot V[i]  \right \} 
\end{matrix}\right.
\label{eq:Max/Min}
\end{equation}

Because render to texture (RTT) only supports 16 texture units in a shader, the whole geometry slices of volumetric datasets cannot be stored and operated using a large number of independent texture units. A practical solution is to rely on a $Texture2DArray$ to save arbitrarily scalable geometry slices. However, this requires a  step to read the rendered image from the frame buffer and write it into the $\Blue{LightBuffer}$. So, the rendered images that are callback from the frame buffer are stored into the light buffer of texture2DArray by the slice index. 

\subsubsection{Compute volume shadow} 
The volume shadow of volumetric datasets can enhance the perception of shape and depth. Although the volume illumination can be carried out by estimating the normal at the sampling point and using the illumination model in ray casting or implemented through the slice-based half-angle slicing technique, interactive volume illumination is difficult to calculate accurately through volume equation. 

In this paper, volume shadow is computed efficiently using the illumination attenuation buffer. For each sample of ray casting, the primary color of volumetric datasets is readily available. The problem here is how to use the $\Blue{LightBuffer}$ to determine the illumination attenuation for each sample. First, each sample is required to find which slice of the $\Blue{LightBuffer}$ it belongs to. And then, the sample $UV$ of the slice needs to be calculated.

\begin{figure}[htb]
  \centering
  \includegraphics[width=1.0\linewidth]{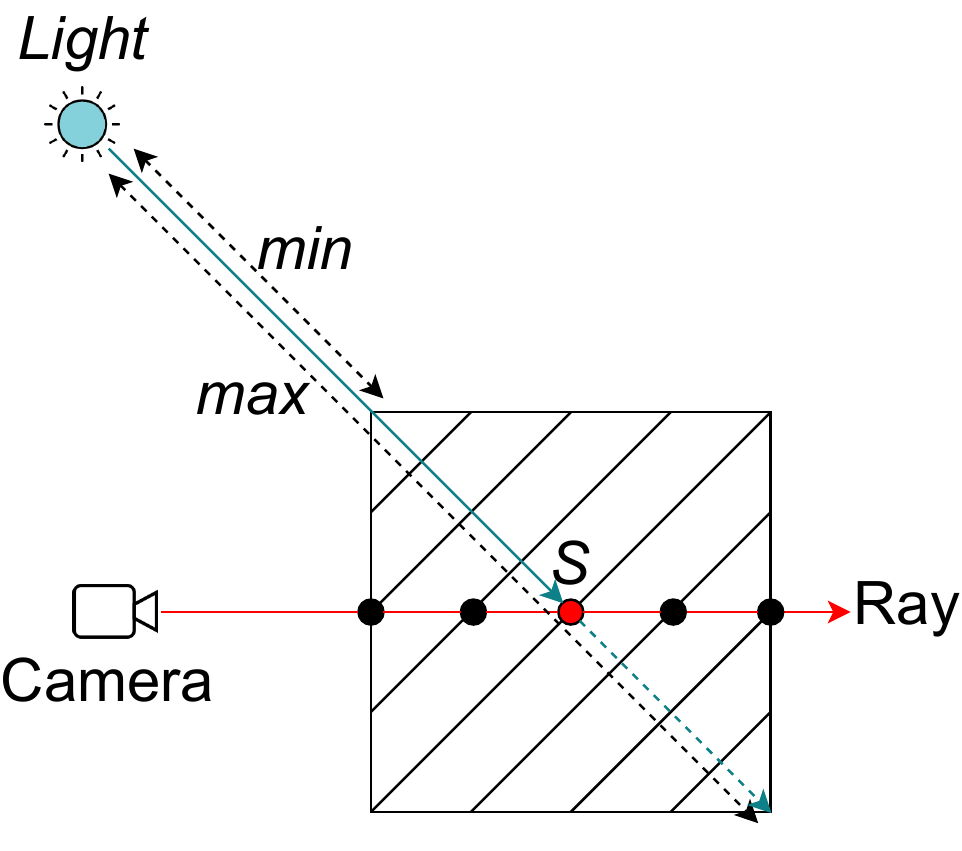}
  %
  \caption{\label{fig:BufferIndex}
           The index of attenuation buffer at a sample. The min/max distance of unit cube vertices are calculated by doing a dot product of each unit cube vertex with the light direction vector.}
\end{figure}

The slice location of the sample is determined by calculating the distance between the sample and the light source. For example (see \textbf{Figure~\ref{fig:BufferIndex}}), the red sample $S$ in ray casting is located in one slice according to the process of building illumination attenuation buffer. Therefore, the solution is calculated by the ratio method, and the index of $2DTextureArray$ is shown as in Equation \ref{eq:BufferIndex}.

\begin{equation}
Index = \frac{n(L \cdot S_{w} - min)}{max - min}
\label{eq:BufferIndex}
\end{equation}
with $n$ is the number of slices, and $S_{w}$ is the world position of sample.

The sample $UV$ (Equation \ref{eq:UV}) of the slice needs to be calculated by the coordination space transform. Similarly to the shadow mapping algorithm, the shadow matrix is used to look up the light buffer, which is performed in the fragment shader. Texture coordinates are computed by matrix transformation in which each vertex position of the object space is multiplied by the shadow matrix to get the shadow texture lookup coordinates. The shadow matrix is multiplied by the model-view projection matrices from the viewpoint of the light. 

\begin{equation}
UV = L_{p} * L_{v} * E_{vi} * E_{mv} * V
\label{eq:UV}
\end{equation}
with $V$ is the object-space vertex position. $L_{p}$ and $L_{v}$ are the projection matric and the view matrix from the viewpoint of the light source. $E_{mv}$ and $E_{vi}$ are the model-view matrix and the view inverse matrix from the viewpoint of the camera.

\subsubsection{Compute scattering coefficient}
\begin{itemize}
\item The shell scattering distribution

The discretized coefficient summation is approximate by a series of cuboid shells, as indicated in \textbf{Figure~\ref{fig:Shell}}, where the number and size of the shells can be varied, for example, $Sh_{0}$, $Sh_{1}$, $Sh_{2}$. The scattering coefficient of the current sample is calculated by estimating each voxel in the shell.

In the test of our algorithm, the current sample moves different sample steps along the $x$, $y$, and $z$ axes to obtain neighboring voxels illumination. Different neighborhoods are given a certain weight to determine the current sample illumination. A more extensive set of shells with varying diameters leads to better image quality but requires more computational overhead; moreover, as few as shells are sufficient to reach an image quality that is hardly distinguishable from individually sampling a large neighborhood.

\begin{figure}[htb]
  \centering
  \includegraphics[width=0.9\linewidth]{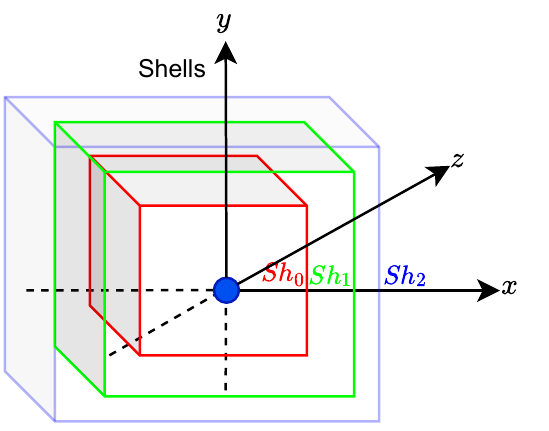}
  %
  \caption{\label{fig:Shell}
           The shell distribution.}
\end{figure}

\item The cone scattering distribution

The cone scattering distributionn is sampled in a cone by the the main light ray as shown in \textbf{Figure~\ref{fig:Cone}(a)}. The cone size is determined by sample length $c_{n}$ at light direction $L_{dir}$ and slice distance $d$ , which is the radius centered on $c_{n}$ in a plane perpendicular to $L_{dir}$, as shown in \textbf{Figure~\ref{fig:Cone}(b)}. Each sample $v$ (as shown Equation \ref{eq:SampleVertex}) in a cone is calculated according axis projection $A_{proj}$ of $c_{n}$ (such as $c_{1}$, $c_{2}$). Meanwhile, the $A_{proj}$ is Rodrigues' ratation formula for rotating the projection base $B_{proj}$ by an angle $\theta$ (as shown Equation \ref{eq:ProjectionAxis}) and the $B_{proj}$ is determined by four points of sample point $s$, light source, the viewer point and projection position $v$ on the same plane.

\begin{figure}[htb]
  \centering
  \includegraphics[width=1.0\linewidth]{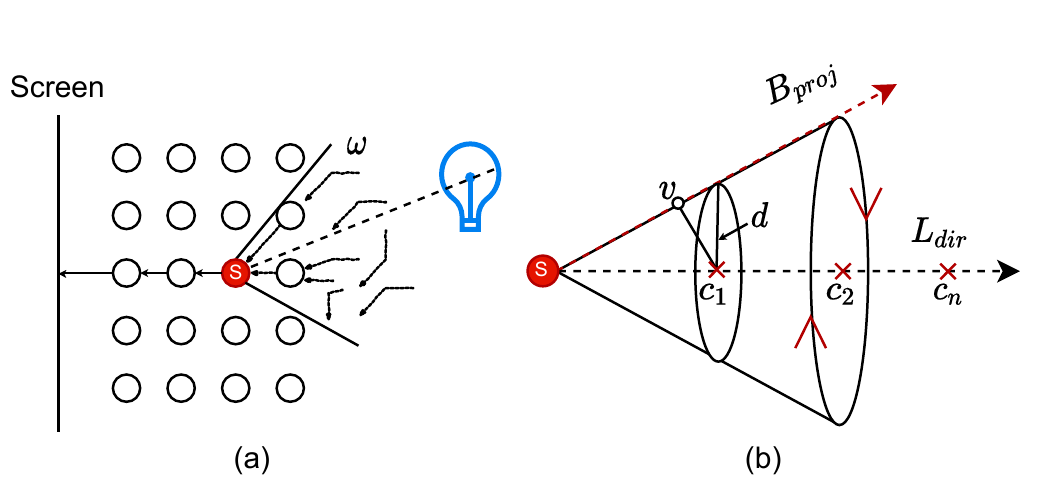}
  %
  \caption{\label{fig:Cone}
           The cone scattering distribution.}
\end{figure}

\begin{equation}
\begin{split}
A_{proj}=B_{proj}\cos \theta + ( L_{dir}\times B_{proj})\sin \theta \\
+( L_{dir} \cdot B_{proj}) L_{dir}(1-\cos \theta)
\label{eq:ProjectionAxis}
\end{split}
\end{equation}

\begin{equation}
v=\Pi _{A_{proj}}(c_{n}) = \frac{c_{n} \cdot A_{proj}} {A_{proj} \cdot A_{proj}} A_{proj}
\label{eq:SampleVertex}
\end{equation}

\end{itemize}

\section{Results and analysis}
To make the proposed algorithm usable cross-platform, version 3.6.4 of OSG (OpenSceneGraph) is selected to implement all the algorithms involved in this paper, and all the tests are performed on different hardware platforms. The related experiments were implemented on a laptop with an NVIDIA GeForce RTX 2070 with Max-Q Design, with 8GB of graphics memory, and an Intel(R) Core(TM) i7-8750H CPU with 16GB of RAM and also on an ordinary Mac with processor 3.2GHz Intel Core i5, memory 16GB 1600MHz, graphics NVIDIA GTX 680MX 2048 MB and the OS X EI Capitan version 10.11.6.

The experimental data of the following tests are from the volume datasets of UZH VMML (\\https://www.ifi.uzh.ch/en/vmml/research/datasets.html) and open scientific visualization datasets (\\https://klacansky.com/open-scivis-datasets/). Furthermore, those files in .raw format of volumetric datasets are loaded from disk and parsed in real-time, and the viewport size of all tests is the same 512x512.

\begin{figure*}[tbp]
  \centering
  \includegraphics[width=1.0\linewidth]{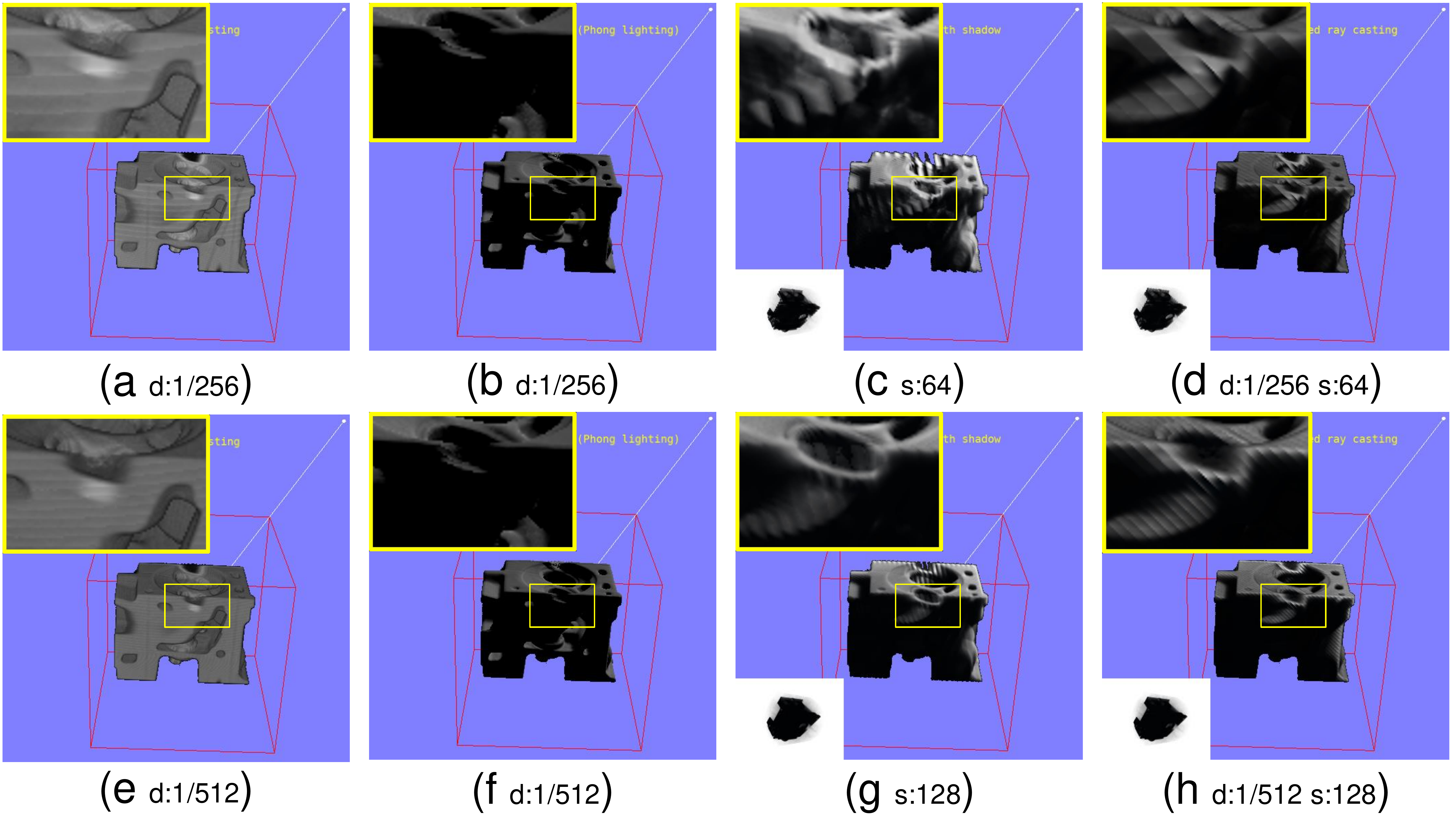}
  \caption{\label{fig:VolumeShadow}%
           \textbf{(a)(e)} Volume rendering by ray casting; \textbf{(b)(f)} Phong lighting volume shadow by ray casting; \textbf{(c)(g)} Volume shadow by half-angle slicing; \textbf{(d)(h)} Volume shadow by slice-based ray casting. The images of \textbf{(c)(d)(g)(h)} at the bottom left are the illumination attenuation of the last slice. `d' is the sample density in ray casting and `s' is the number of slices of the volumetric datasets $Egnine$($256^3$). The red lines are the unit cube boundary. The white point at the top right is the light position.}
\end{figure*}

Volume rendering can be implemented easily using single-pass GPU ray casting with alpha compositing as shown in \textbf{Figure~\ref{fig:VolumeShadow} (a)(e)}. The obtained samples during the ray traversal in ray casting are composited using the front-to-back alpha compositing until the ray exits the volumetric datasets. Meanwhile, the normal at the sample point is estimated to carry out the local illumination calculation using the Phong illumination model, as shown in \textbf{Figure~\ref{fig:VolumeShadow} (b)(f)}. Although the local volume illumination is adequate for shading surfaces, it does not provide sufficient lighting characteristics for the visual appearance. The global volume illumination, half-angle slicing for volume shadow (\textbf{Figure~\ref{fig:VolumeShadow} (c)(g)}) and our slice-based ray casting (\textbf{Figure~\ref{fig:VolumeShadow} (d)(h)}) presents detailed lighting characteristics in contrast to the local volume illumination. For example, the global illumination can see the structure on the wall of the hole, and slice-based ray casting can have a clearer appearance, especially when fewer slices are compared to the half-angle slicing.

\begin{figure}[tbp]
  \centering
  \includegraphics[width=1.0\linewidth]{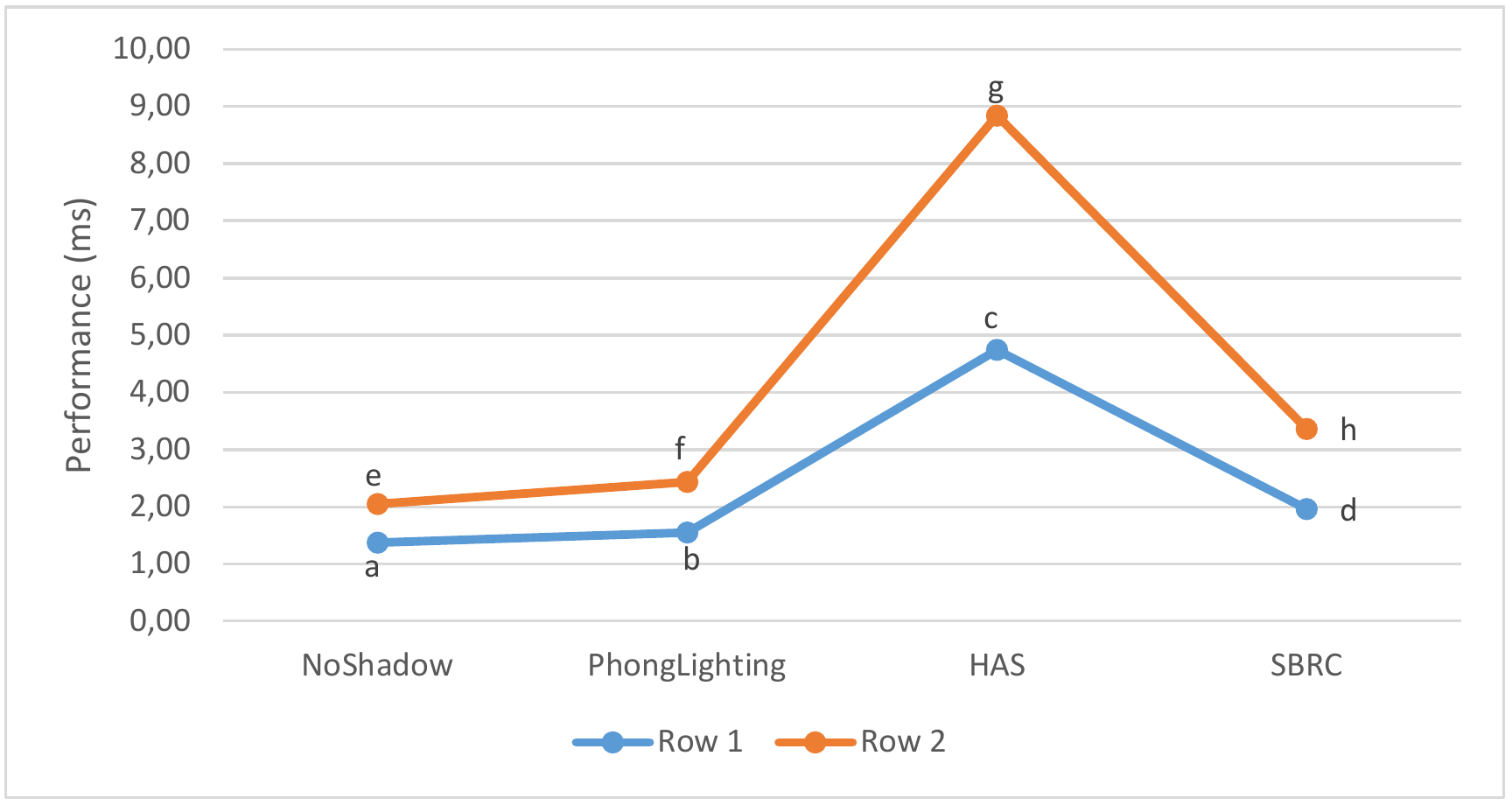}
  \caption{\label{fig:VolumeShadowPerformance}%
           The performance contrast of four methods (no shadow, Phong lighting, half-angle slicing (HAS) and slice-based ray casting (SBRC)) in \textbf{Figure~7} and all the performance (ms) are averages over time. Row 1 is a, b, c, d. Row 2 is e, f, g, h.}
\end{figure}

In volume ray casting rendering, as the sample density increases, the results are more fined less artifact. The sample density $1/256$ and $1/512$ are one unit and half unit size of the volumetric dataset $Engine (256^3)$, respectively. It also can be seen from the \textbf{Figure~\ref{fig:VolumeShadow}} that as the number of slices increases, the effects of both half-angle slicing and slice-based ray casting are better for the illumination effects. However, the half-angle slicing requires a large number of passes during the whole volume illumination rendering; thus, the draw calls are huge to result in high-performance overhead, as shown in \textbf{Figure~\ref{fig:VolumeShadowPerformance}}. Our slice-based ray casting can supply high-performance volume illumination rendering. It only increases the computing of initial building illumination buffer and per sample shadow computing in ray casting. And the slice-based ray casting will not have much performance fluctuation as the number of slices increases.

\begin{figure}[tbp]
  \centering
  \includegraphics[width=1.0\linewidth]{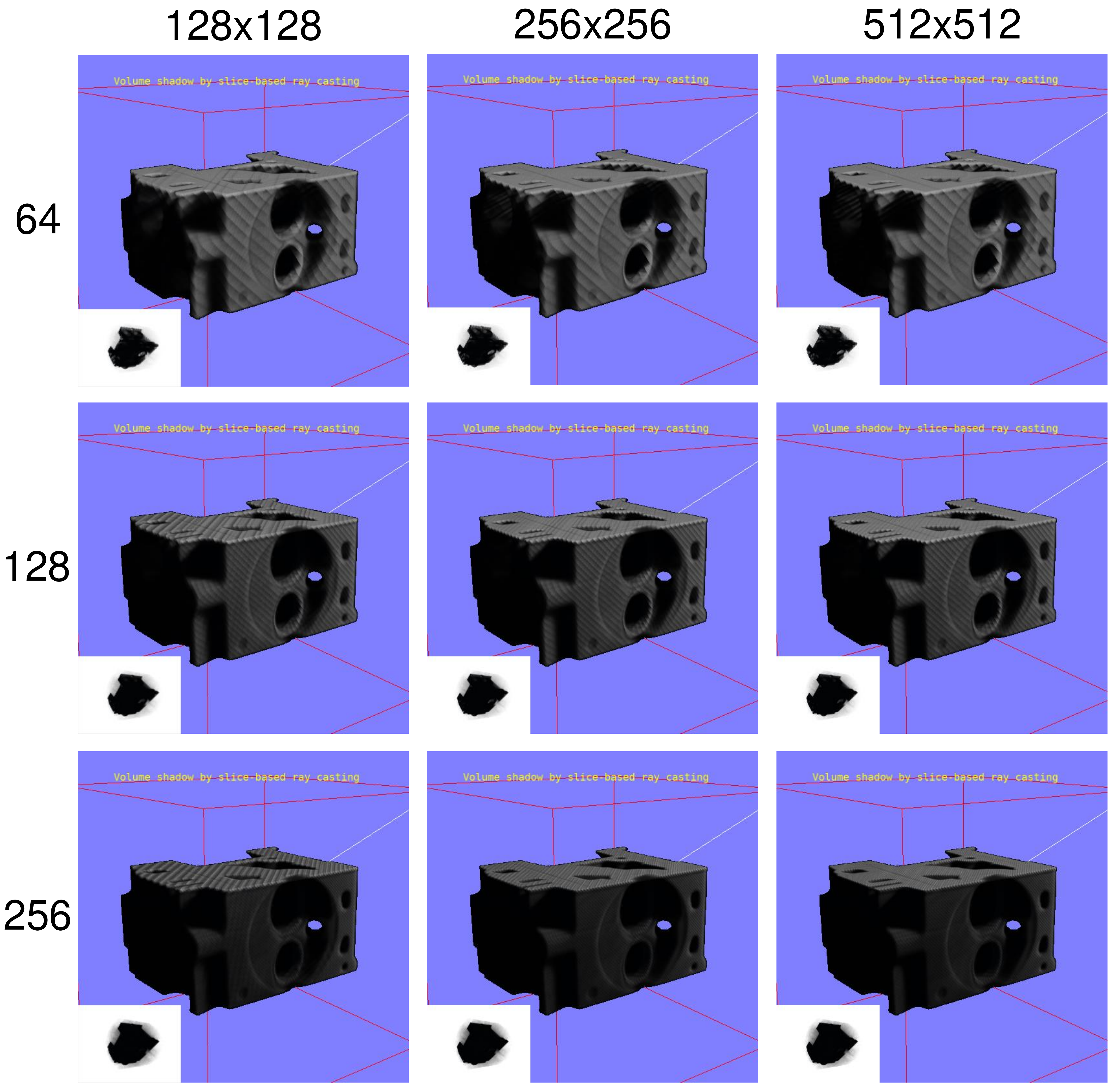}
  \caption{\label{fig:ScalablePerformance}%
           The top is three resolution of 128x128, 256x256, and 512x512. The left is the number of slices of 64, 128, and 256. The images of at the bottom left are the illumination attenuation of the last slice.}
\end{figure}

\begin{figure}[tbp]
  \centering
  \includegraphics[width=1.0\linewidth]{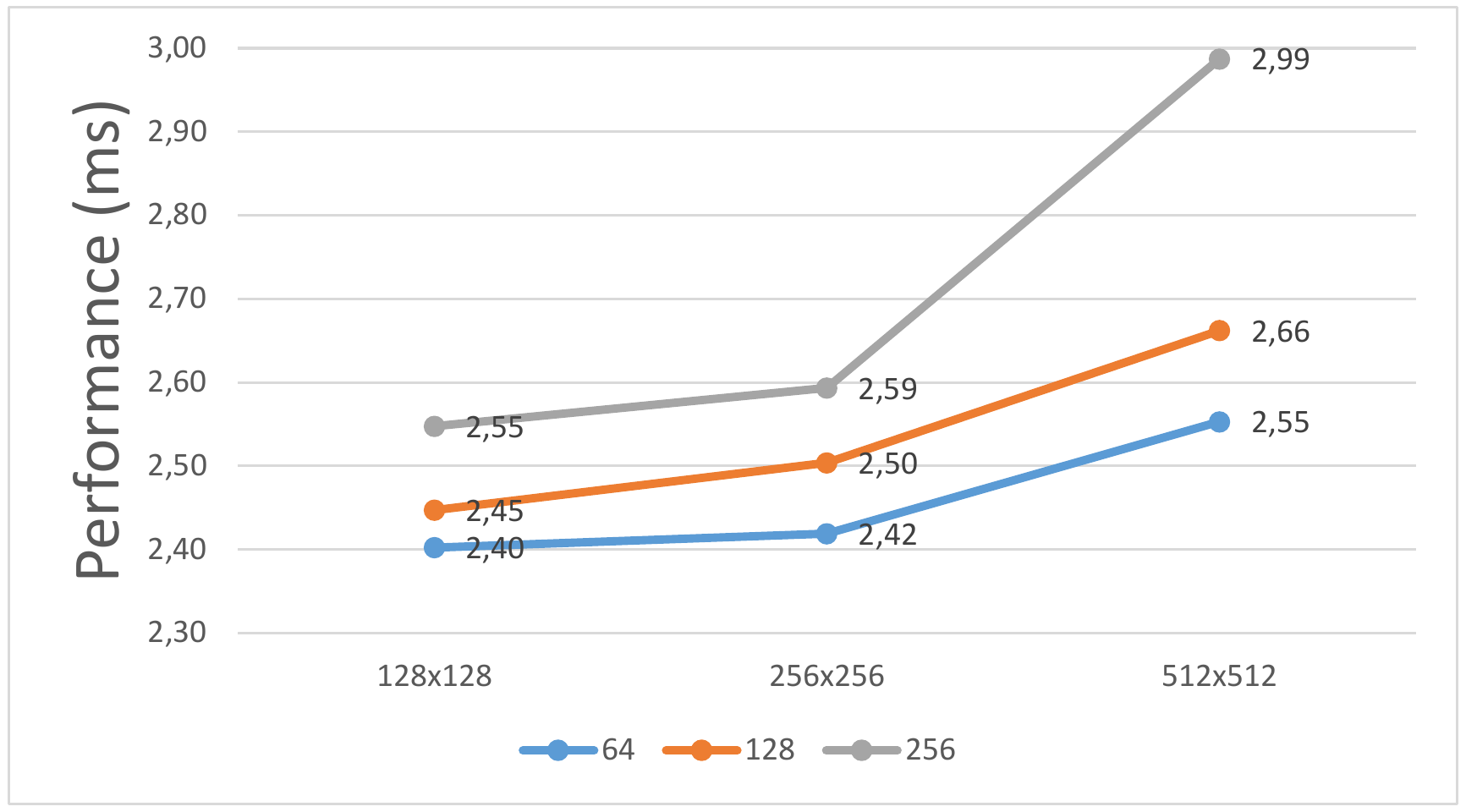}
  \caption{\label{fig:ScalablePerformanceChart}%
           The flexible performance contrast in three different resoution of 128x128, 156x256 and 512x512 and three the number of slices of 64, 128, 256 in \textbf{Figure~8}.}
\end{figure}

Moreover, our slice-based ray casting allows for flexible performance, varying the number of slices and per-slice image resolution \textbf{Figure~\ref{fig:ScalablePerformanceChart}}. As the number of slices and image resolution increases, the effects of volume illumination are better, but the performance overhead is increased, as shown in \textbf{Figure~\ref{fig:ScalablePerformance}}. In the test, if the number of slices exceeds 256 and the image resolution exceeds 512x512, the performance will decrease sharply. So, the parameters of slice-based ray casting should be chosen to meet flexible performance according to practical applications and hardware performance.

\begin{figure*}[tbp]
  \centering
  \includegraphics[width=1.0\linewidth]{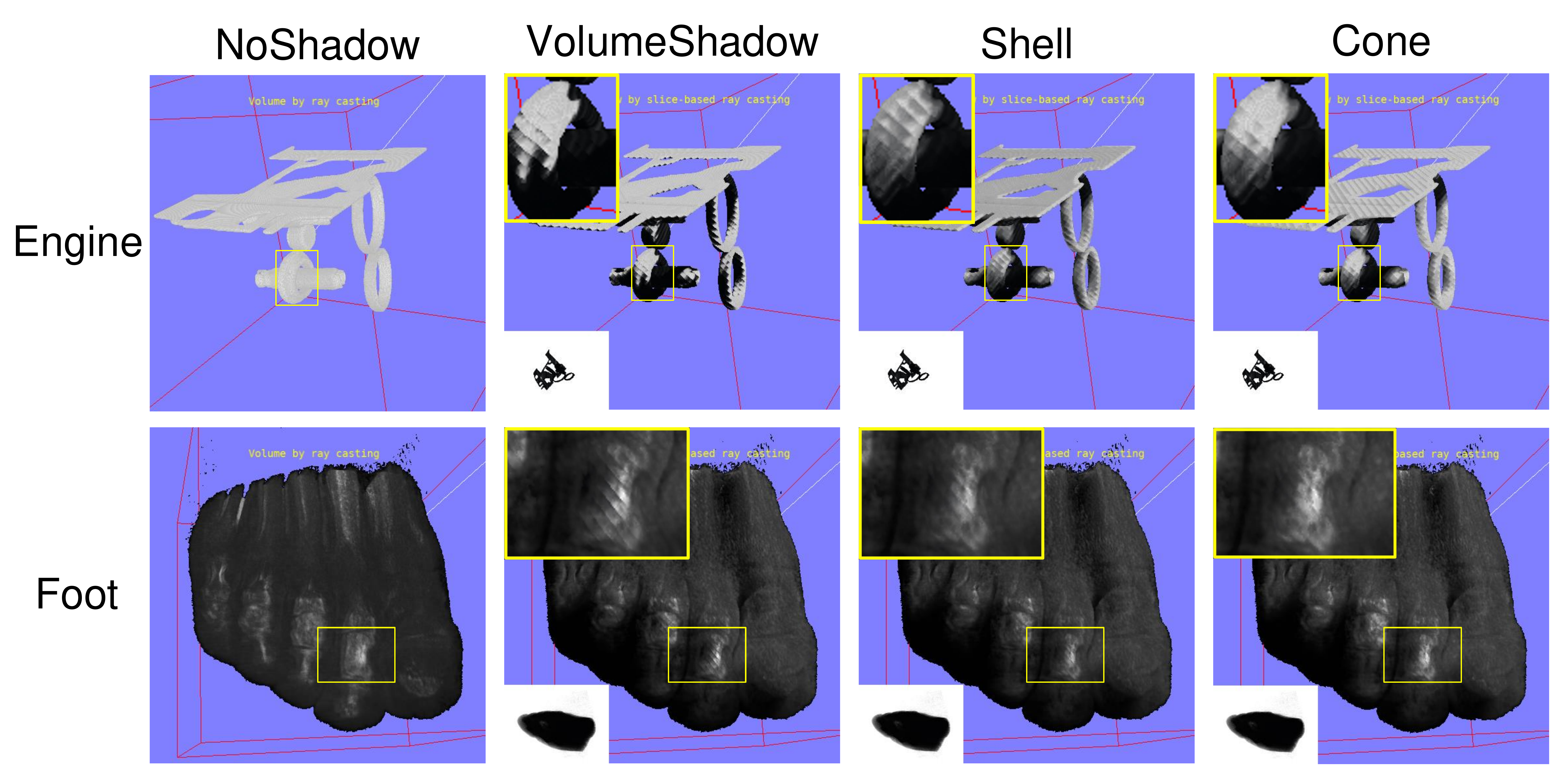}
  \caption{\label{fig:ShellCone}%
           The contrast of volume illumination for the shell and cone sample distribution. The engine and foot are the sizes $256^3$ and the number of slices of illumination is 128. The images of at the bottom left are the illumination attenuation of the last slice.}
\end{figure*}

\begin{figure}[tbp]
  \centering
  \includegraphics[width=1.0\linewidth]{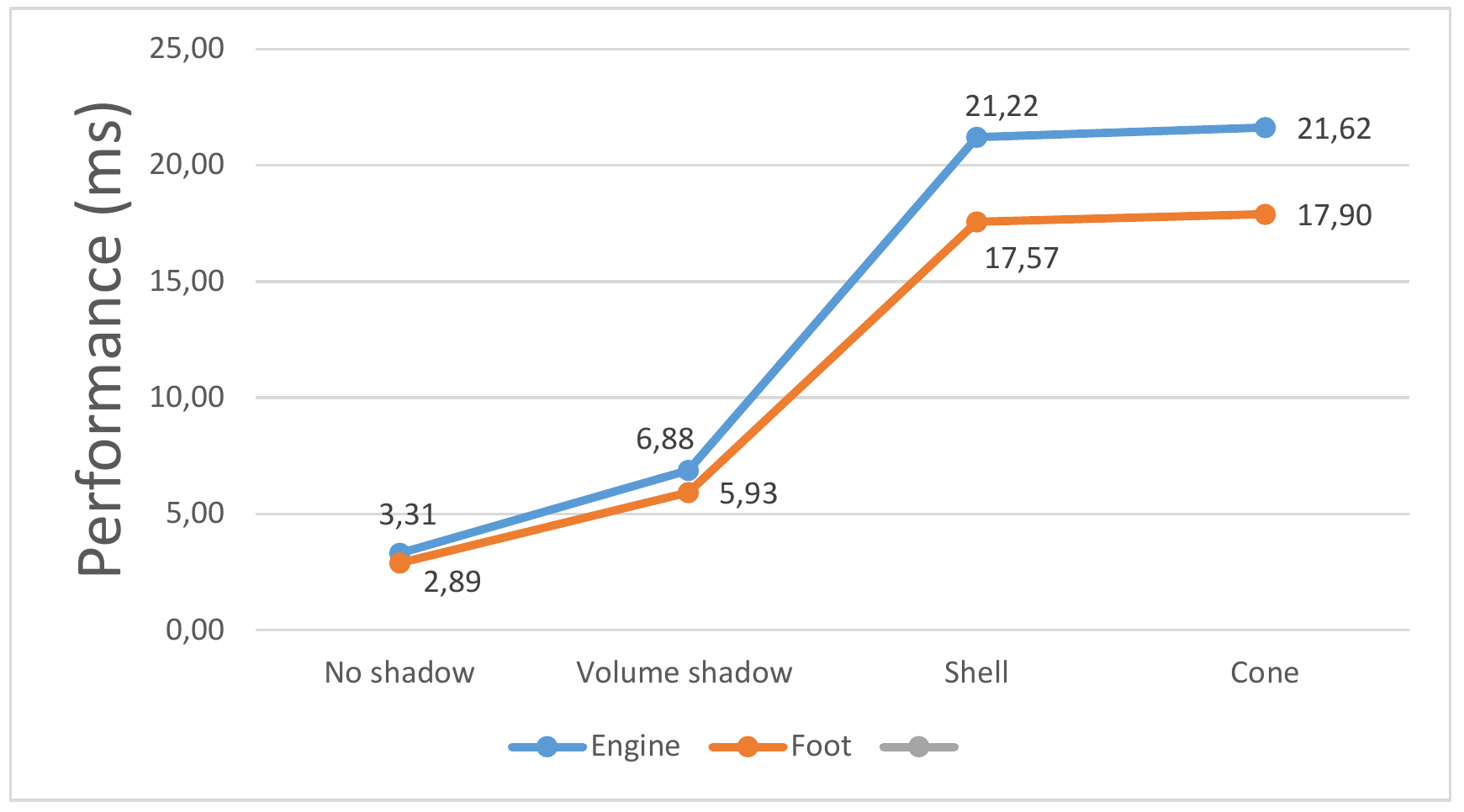}
  \caption{\label{fig:ShellConePerformance}%
           The performance contrast in \textbf{Figure~8}.}
\end{figure}

In addition to slice-based ray casting, which can quickly calculate volume shadow illumination, the illumination scattering effects can be expressed by approximating the volume illumination equation. The shell and cone sample distribution be used to generate the soft shadow, as shown in \textbf{Figure~\ref{fig:ShellCone}}. The cone effects are better than shell because the cone is integrated with the light direction rather than the surround voxels of the sample point in the shell. Meanwhile, cone and shell performance is related to the number of samples. As shown in \textbf{Figure~\ref{fig:ShellConePerformance}}, the shell uses six neighborhood sample points, and the cone is eight sample points. So the cone performance is slightly higher than the shell, but the specular illumination much better.

In these tests, some image effects are somewhat dark, which is related to the cumulative accumulation of illumination attenuation. Therefore, as the number of slices increases, multiplying each slice by $(1 + \alpha)^n$ (The $\alpha$ is an alpha value of the current slice, and the $n$ is an adjustable factor of illumination attenuation) compensates for the fact that more and more slices are over-dark due to excessive accumulation of illumination attenuation.

\section{Conclusions and future work}

In short, the paper proposes a flexible performance volume illumination rendering that combines illumination attenuation slice from the light source and effective scattering coefficients in ray casting. The method can quickly obtain volume shadows and better illumination effects in contrast to the local volume illumination or other similar slice-based global volume illumination. In comparison to different slice-based volume illumination algorithms such as half-angle slicing, slice-based ray casting can improve the rendering performance for volume illumination. Meanwhile, the scalability of slice-based ray casting can also be flexibly applied to more volumetric datasets and interactive volume rendering applications. However, there are also some problems with slice-based ray casting to represent the better volume illumination effects, such as the construction of better illumination attenuation slices and the calculation of better shell and cone scattering coefficients. 

In the future, we will further optimize and improve the slice-based ray casting method to form a more sophisticated and more valuable volume illumination method. Meanwhile, combined multi-slice per pass, which reduces the performance overhead of building illumination attenuation buffer, provides a more flexible and scalable framework of volume illumination rendering.

\bibliographystyle{unsrt}  
\bibliography{references}  

\begin{thebibliography}{10}

\bibitem{beyer2018gpu}
Johanna Beyer and Markus Hadwiger.
\newblock Gpu-based large-scale scientific visualization.
\newblock In {\em SIGGRAPH Asia 2018 Courses}, pages 1--217. 2018.

\bibitem{MK:16}
Finian Mwalongo, Michael Krone, Guido Reina, and Thomas Ertl.
\newblock State-of-the-art report in web-based visualization.
\newblock In {\em Computer graphics forum}, volume~35, pages 553--575. Wiley
  Online Library, 2016.

\bibitem{hachaj2014real}
Tomasz Hachaj.
\newblock Real time exploration and management of large medical volumetric
  datasets on small mobile devices—evaluation of remote volume rendering
  approach.
\newblock {\em International Journal of Information Management},
  34(3):336--343, 2014.

\bibitem{noguera2015mobile}
Jose~M Noguera and J~Roberto Jimenez.
\newblock Mobile volume rendering: past, present and future.
\newblock {\em IEEE transactions on visualization and computer graphics},
  22(2):1164--1178, 2015.

\bibitem{HW:16}
Claudia H{\"a}nel, Benjamin Weyers, Bernd Hentschel, and Torsten~W Kuhlen.
\newblock Visual quality adjustment for volume rendering in a head-tracked
  virtual environment.
\newblock {\em IEEE transactions on visualization and computer graphics},
  22(4):1472--1481, 2016.

\bibitem{JS:14}
Daniel J{\"o}nsson, Erik Sund{\'e}n, Anders Ynnerman, and Timo Ropinski.
\newblock A survey of volumetric illumination techniques for interactive volume
  rendering.
\newblock In {\em Computer Graphics Forum}, volume~33, pages 27--51. Wiley
  Online Library, 2014.

\bibitem{EM:19}
M~El~Seoud and Amr~S Mady.
\newblock A comprehensive review on volume rendering techniques.
\newblock In {\em Proceedings of the 2019 8th International Conference on
  Software and Information Engineering}, pages 126--131. ACM, 2019.

\bibitem{BS:05}
Udeepta~D Bordoloi and H-W Shen.
\newblock View selection for volume rendering.
\newblock In {\em VIS 05. IEEE Visualization, 2005.}, pages 487--494. IEEE,
  2005.

\bibitem{Milan}
Ikits Milan, Joe Kniss, Lefohn Aaron, and Hansen Charles.
\newblock Chapter 39. volume rendering techniques.
\newblock [EB/OL].
\newblock
  \url{https://developer.download.nvidia.com/books/HTML/gpugems/gpugems_ch39.html}.

\bibitem{kruger2003acceleration}
Jens Kruger and R{\"u}diger Westermann.
\newblock Acceleration techniques for gpu-based volume rendering.
\newblock In {\em IEEE Visualization, 2003. VIS 2003.}, pages 287--292. IEEE,
  2003.

\bibitem{SSK:05}
Simon Stegmaier, Magnus Strengert, and Thomas Klein.
\newblock A simple and flexible volume rendering framework for
  graphics-hardware-based raycasting.
\newblock In {\em Proceedings Conference on Volume Graphics}, pages 187--195,
  2005.

\bibitem{LL:94}
Philippe Lacroute and Marc Levoy.
\newblock Fast volume rendering using a shear-warp factorization of the viewing
  transformation.
\newblock In {\em Proceedings ACM SIGGRAPH}, pages 451--458, 1994.

\bibitem{LH:91}
David Laur and Pat Hanrahan.
\newblock Hierarchical splatting: A progressive refinement algorithm for volume
  rendering.
\newblock In {\em Proceedings ACM SIGGRAPH}, pages 285--288. ACM Press, 1991.

\bibitem{EKE:01}
Klaus Engel, Martin Kraus, and Thomas Ertl.
\newblock High-quality pre-integrated volume rendering using
  hardware-accelerated pixel shading.
\newblock In {\em Proceedings ACM SIGGRAPH/EUROGRAPHICS Workshop on Graphics
  Hardware}, pages 9--16, 2001.

\bibitem{FR:04}
Randima Fernando et~al.
\newblock {\em {GPU} gems: programming techniques, tips, and tricks for
  real-time graphics}, volume 590.
\newblock Addison-Wesley Reading, 2004.

\bibitem{KPHE:02}
Joe Kniss, Simon Premoze, Charles Hansen, and David Ebert.
\newblock Interactive translucent volume rendering and procedural modeling.
\newblock In {\em Proceedings IEEE Visualization}, pages 109--116, 2002.

\bibitem{EHKRW:06}
Klaus Engel, Markus Hadwiger, Joe~M. Kniss, Christof Rezk-Salama, and Daniel
  Weiskopf.
\newblock {\em Real-Time Volume Graphics}.
\newblock AK Peters, 2006.

\bibitem{ZE:11}
Qi~Zhang, Roy Eagleson, and Terry~M Peters.
\newblock Volume visualization: a technical overview with a focus on medical
  applications.
\newblock {\em Journal of digital imaging}, 24(4):640--664, 2011.

\bibitem{SMP:11}
Philipp Schlegel, Maxim Makhinya, and Renato Pajarola.
\newblock Extinction-based shading and illumination in {GPU} volume
  ray-casting.
\newblock {\em IEEE Transactions on Visualization and Computer Graphics},
  17(12):1795--1802, December 2011.

\bibitem{zhang2013lighting}
Yubo Zhang and Kwan-Liu Ma.
\newblock Lighting design for globally illuminated volume rendering.
\newblock {\em IEEE transactions on visualization and computer graphics},
  19(12):2946--2955, 2013.

\bibitem{ZY:13}
Yubo Zhang and Kwan-Liu Ma.
\newblock Lighting design for globally illuminated volume rendering.
\newblock {\em IEEE Transactions on Visualization and Computer Graphics},
  19(12):2946--2955, 2013.

\bibitem{PB:16}
Bernhard Preim, Alexandra Baer, Douglas Cunningham, Tobias Isenberg, and Timo
  Ropinski.
\newblock A survey of perceptually motivated 3d visualization of medical image
  data.
\newblock In {\em Computer Graphics Forum}, volume~35, pages 501--525. Wiley
  Online Library, 2016.

\bibitem{schott2009directional}
Mathias Schott, Vincent Pegoraro, Charles Hansen, K{\'e}vin Boulanger, and Kadi
  Bouatouch.
\newblock A directional occlusion shading model for interactive direct volume
  rendering.
\newblock In {\em Computer Graphics Forum}, volume~28, pages 855--862. Wiley
  Online Library, 2009.

\bibitem{vsolteszova2010multidirectional}
Veronika {\v{S}}olt{\'e}szov{\'a}, Daniel Patel, Stefan Bruckner, and Ivan
  Viola.
\newblock A multidirectional occlusion shading model for direct volume
  rendering.
\newblock In {\em Computer Graphics Forum}, volume~29, pages 883--891. Wiley
  Online Library, 2010.

\bibitem{patel2013instant}
Daniel Patel, Veronika {\v{S}}olt{\'e}szov{\'a}, Jan~Martin Nordbotten, and
  Stefan Bruckner.
\newblock Instant convolution shadows for volumetric detail mapping.
\newblock {\em ACM Transactions on Graphics (TOG)}, 32(5):1--18, 2013.

\bibitem{LKGHHY:16}
Patric Ljung, Jens Kr{\"u}ger, Eduard Gr{\"o}ller, Markus Hadwiger, Charles~D.
  Hansen, and Anders Ynnerman.
\newblock State of the art in transfer functions for direct volume rendering.
\newblock {\em Computer Graphics Forum}, 35(3):669--691, June 2016.

\bibitem{ME:17}
Bo~Ma and Alireza Entezari.
\newblock Volumetric feature-based classification and visibility analysis for
  transfer function design.
\newblock {\em IEEE transactions on visualization and computer graphics},
  24(12):3253--3267, 2017.

\bibitem{UJ:17}
Will Usher, Jefferson Amstutz, Carson Brownlee, Aaron Knoll, and Ingo Wald.
\newblock Progressive cpu volume rendering with sample accumulation.
\newblock In {\em Proceedings of the 17th Eurographics Symposium on Parallel
  Graphics and Visualization}, pages 21--30. Eurographics Association, 2017.

\bibitem{BHP:15}
Johanna Beyer, Markus Hadwiger, and Hanspeter Pfister.
\newblock State-of-the-art in {GPU}-based large-scale volume visualization.
\newblock {\em Computer Graphics Forum}, 34(8):13--37, December 2015.

\bibitem{WC:19}
Sebastian Weiss, Mengyu Chu, Nils Thuerey, and R{\"u}diger Westermann.
\newblock Volumetric isosurface rendering with deep learning-based
  super-resolution.
\newblock {\em arXiv preprint arXiv:1906.06520}, 2019.

\bibitem{rostamzadeh2013comparison}
Neda Rostamzadeh, Daniel J{\"o}nsson, and Timo Ropinski.
\newblock A comparison of volumetric illumination methods by considering their
  underlying mathematical models.
\newblock In {\em Proceedings of SIGRAD 2013; Visual Computing; June 13-14;
  2013; Norrk{\"o}ping; Sweden}, number 094, pages 35--40. Link{\"o}ping
  University Electronic Press, 2013.

\bibitem{max1995optical}
Nelson Max.
\newblock Optical models for direct volume rendering.
\newblock {\em IEEE Transactions on Visualization and Computer Graphics},
  1(2):99--108, 1995.

\end{thebibliography}

\end{document}